\input harvmac.tex
\Title{\vbox{\baselineskip12pt\hbox{MRI-PHY/13/95}\hbox{CTP-TAMU-53/95}
\hbox{hep-th/9512157}}}
{\vbox{\centerline{Dyonic Black Hole in Heterotic String Theory}}}

\centerline{\bf Dileep P. Jatkar$^1$, Sudipta Mukherji$^2$ and Sudhakar
Panda$^1$}
\smallskip\centerline{\it $^1$Mehta Research Institute of Mathematics \&
Mathematical Physics}
\centerline{\it 10, Kasturba Gandhi Marg, Allahabad 211 002, India}
\centerline{\it $^2$Center for Theoretical Physics, Department of Physics}
\centerline{\it Texas A\& M University, College Station, Texas 77843-4242, USA}

\vskip .3in
We study some features of dyonic Black hole solution in heterotic string
theory on a six torus. This solution has 58 parameters. Of these, 28 parameters
denote the electric charge of the black hole, another 28 correspond to the
magnetic charge, and the other two parameters being the mass and the
angular momentum of the black hole. We discuss the extremal limit and show
that in various limits, it reduces to the known black hole solutions.  The
solutions saturating the Bogomolnyi bound are identified. 
Explicit solution is presented for the non-rotating dyonic black hole.

\Date{12/95}

\def \sa {\sinh\alpha} \def \sb {\sinh\beta} \def \sg {\sinh\gamma}
\def \sd {\sinh\delta} \def \ca {\cosh\alpha} \def \cb {\cosh\beta}
\def \cg {\cosh\gamma} \def \cd {\cosh\delta} 
 \def \cu {\cos{u}} \def \su {\sin{u}}
\def \sta {\sinh^2\alpha} \def \stb {\sinh^2\beta}
\def \stg {\sinh^2\gamma} \def \std {\sinh^2\delta} \def \cta {\cosh^2\alpha}
\def \ctb {\cosh^2\beta} \def \ctg {\cosh^2\gamma} \def \ctd {\cosh^2\delta}
  
 \def \cT {\cos{T}} \def \sT {\sin{T}}
\def \cR {\cos{R}} \def \sR {\sin{R}} \def \ctT {\cos^2{T}}
\def \stT {\sin^2{T}} \def \ctR {\cos^2{R}} \def \stR {\sin^2{R}}
 \def \ccb {\cosh^3\beta} 
\def \ccd {\cosh^3\delta} \def \cfa {\cosh^4\alpha} 
\def \cfg {\cosh^4\gamma} 
\def \ad {a_{\Delta}} \def \bd {b_{\Delta}} \def \r {\rho}
\lref\chsone{C. G. Callan, J. A. Harvey and A. Strominger, Nucl. Phys.
{\bf B359} (1991) 611.}

\lref\strsol{M. J. Duff, R. R. Khuri and J. X. Lu, {\it String Solitons},
hep-th/9412184 (submitted to Phys. Rep.).}

\lref\sen {A. Sen {\it Black Hole Solutions in Heterotic String Theory on a
Torus}, TIFR-TH-94-47, hep-th/9411187}

\lref\sendual{A. Sen {\it Strong-Weak Coupling Duality in the Four
Dimensional Heterotic String Theory, hep-th/9402002}}

\lref\asen {A. Sen, Phys. Lett. {\bf B271} (1991) 295.}

\lref\duff{M. J. Duff and J. Rahmfeld, Phys. Lett. {\bf B345} (1995) 441.}

\lref\senbh {A. Sen {\it Extremal Black Holes and Elementary String States},
hep-th/9504147}

\lref\hasen {S. F. Hassan and A. Sen, {\it Twisting Classical Solutions in
Heterotic String Theory}, TIFR-TH-94-47, hep-th/9109038}

\lref\chstwo{C. G. Callan, J. A. Harvey and A. Strominger, Nucl. Phys.
{\bf B367} (1991) 60.}

\lref\dghrr{A. Dabholkar, G. Gibbons, J. A. Harvey and F. Ruiz Ruiz,
Nucl. Phys. {\bf B340} (1990) 33.}

\lref\dabhar{A. Dabholkar and J. A. Harvey, Phys. Rev. Lett. {\bf 63}
(1989) 478.}

\lref\mandal{G. Mandal and S. R. Wadia, {\it Black Hole Geometry around an
elementary BPS String State}, hep-th/9511218.}

\lref\callan{C. G. Callan, J. M. Maldacena and A. W. Peet, {\it Extremal
Black Holes as Fundamental Strings}, hep-th/9510134.}

\lref\dghw{A. Dabholkar, J. Gauntlett, J. Harvey and D. Waldram,
hep-th/9511053.}

\lref\kallosh{R. Kallosh, D Kastor, T. Ortin and T. Torma {\it Supersymmetric
and Stationary Solutions in Dilaton Axion Gravity}, hep-th/9406059}

\lref\bhpapers{ A. Sen, {\it Black Holes
and Solitons in String Theory}, hep-th/9210050; G. Horowitz, {\it The
Dark Side of String Theory: Black Holes and Black Strings},
hep-th/9210119; R. R. Khuri, {\it Black Holes and Solitons in String
Theory}, hep-th/9506065.}

\lref\jmp{D. P. Jatkar, S. Mukherji and S. Panda, in preparation.}

\lref\cybps{M Cvetic and D. Youm, {\it Dyonic BPS Saturated Black Holes of
Heterotic string on a Six-torus}, hep-th/9507090.}

\lref\cvetic{M. Cvetic and D. Youm, {\it
All the static spherically symmetric Black Holes of Heterotic String on a
Six Torus}, IASSNS-HEP-95/107, hep-th/9512127.}

\lref\rk{R. Kallosh, private communication.}

\lref\ashoke{A. Sen. {\it Strong-Weak Coupling Duality in Three 
Dimensional String Theory}, hep-th/9408083.}

\newsec{Introduction}

In recent years a variety of approximate and exact solitonic solutions to the
string theory were obtained\refs{\dghrr,\dabhar,\chsone}. For a comprehensive
review of the subject see ref.\refs{\strsol}. The study of soliton solutions
is useful for understanding the non-perturbative aspects of the string
theory, as it has become abundantly clear from the recent progress in the S
and U duality symmetries of the string theory in various dimensions. Black
hole solutions are solitonic solutions of the string theory and they play an
important role in the duality symmetries of the string theory. From the point
of view of the S-duality, which in general relates the perturbative
spectrum
of one string theory to the non-perturbative spectrum of another string
theory, it is therefore important to understand the qualitative features of
solitonic solutions in string theory. For example, in case of black holes it
is known that the entropy of the black hole is proportional to the area of
the horizon. Using the relation of the area of the horizon and the mass of
the black hole one finds that the entropy of the black hole depends
quadratically on its mass. The density of string states at a given mass level
is proportional to the exponential of the mass and therefore the entropy is a
linear function of the mass. Since black holes are also part of the string
spectrum it is desirable to understand the relation between the massive
string states and the black hole\refs{\duff,\senbh}. Explicit form of the
black hole solution may be useful in addressing some of these questions.
Recently there has been a lot of progress in understanding the relation
between extremal black holes and the elementary string states\refs{\callan,
\dghw,\mandal}. Black hole solution to the string equations of motion has
been discussed by several groups in the past\refs{\bhpapers}. Most general
electrically charged black hole solution in the heterotic string theory
was discussed by Sen\refs{\sen}.

In this paper we present the general dyonic black hole solution to the
heterotic string theory. We obtain the solution using the technique
outlined in ref.\refs{\sen}. The plan of the paper is as follows. Section
2 contains a short exposition to the duality transformations\refs{\sen}
within the context of the Kerr solution. Section 3 contains the derivation
of the dyonic black hole solution starting from the Kerr metric and
discussion of the characteristics of the new solution. In section 4 we
discuss various limits of the solution. We  show that in certain limits we
get the extremal black hole solution which saturates the Bogomolnyi bound. 
We also discuss various limits in which this solution
reduces to pure electrically charged, pure magnetically charged as well as
neutral black hole solution. In section 5 we discuss the non-rotating
black hole in detail.

\newsec{Duality Transformations}

We wish to study a general black hole solution to the heterotic string
compactified on a six dimensional torus. The spectrum of massless fields
for a generic six torus compactification of the heterotic string contains
the metric $G_{\mu\nu}$, the antisymmetric tensor $B_{\mu\nu}$, the dilaton
$\Phi$, twenty eight abelian gauge fields $A_{\mu}^{(a)}$ and a $28\times 28$
matrix scalar field $M$ which contains the scalar fields coming from the
internal components of the ten dimensional metric, antisymmetric tensor and
gauge fields. The matrix valued scalar field $M$ satisfies following relations
\eqn\ortho{ MLM^T = L\qquad M^T=M.}
Here $L$ is a $28\times 28$ symmetric matrix and we choose it to be
\eqn\lmet{L=\pmatrix{-I_{22}&\quad\cr\quad&I_6}}
where and in the rest of the paper $I_n$ denotes $n \times n$ identity matrix.
The low energy effective field theory of these fields is given by
\eqn\action{\eqalign{S &= \int d^4x\sqrt{-\det G}\, e^{-\Phi}[R_G+G^{\mu\nu}
\partial_{\mu}\Phi\partial_{\nu}\Phi+{1\over 8}G^{\mu\nu}Tr(\partial_{\mu}ML
\partial_{\nu}ML)\cr
&-{1\over 12}G^{\mu\mu'}G^{\nu\nu'}G^{\rho\rho'}H_{\mu\nu\rho}H_{\mu'\nu'
\rho'}-G^{\mu\mu'}G^{\nu\nu'}F^{(a)}_{\mu\nu}(LML)_{ab}F^{(b)}_{\mu'\nu'}]},}
where,
\eqn\fst{F^{(a)}_{\mu\nu}=\partial_{\mu}A^{(a)}_{\nu}-\partial_{\nu}
A^{(a)}_{\mu},}
\eqn\asfs{H_{\mu\nu\rho}=\partial_{\mu}B_{\nu\rho}+2A^{(a)}_{\mu}L_{ab}
F^{(b)}_{\nu\rho}+ {\rm c.p. in} \mu, \nu {\rm and} \rho,}
and $R_G$ is the Ricci scalar corresponding to the metric $G_{\mu\nu}$.
This action is invariant under an $O(6,22)$ transformations $\Omega$ which
acts on the fields as follows
\eqn\odd{M\rightarrow\Omega M\Omega^T,\quad A^{(a)}_{\mu}\rightarrow
\Omega_{ab}A^{(b)}_{\mu},\quad \Omega L\Omega^T = L,}
and leaves the four dimensional metric $G_{\mu\nu}$, the antisymmetric tensor
$B_{\mu\nu}$ and the dilaton $\Phi$ invariant.

We intend to obtain the general black hole solution using the solution
generating technique\refs{\asen,\hasen} which uses the global symmetries
of the effective field theory to generate new solutions of the equation of
motion from the known solution. This technique has been used extensively to
get various new solutions to the string equations of motion
For definiteness let us start with the Kerr solution
\eqn\kerr{\eqalign{ds^2 &\equiv G_{\mu\nu} dx^\mu dx^\nu\cr
&= -{\rho^2 + a^2 \cos^2\theta - 2m\rho\over \rho^2 + a^2 \cos^2 \theta}
dt^2 +{\rho^2 + a^2 \cos^2 \theta \over \rho^2 + a^2 -2m \rho} d\rho^2
+ (\rho^2 + a^2 \cos^2\theta) d\theta^2 \cr
& + {\sin^2\theta\over \rho^2 + a^2 \cos^2\theta} [(\rho^2+a^2) ( \rho^2
+ a^2 \cos^2\theta) + 2m\rho a^2\sin^2\theta] d\phi^2 \cr
& -{4m\rho a \sin^2\theta \over \rho^2+a^2\cos^2\theta} dt d\phi\cr
&\Phi = 0,\qquad B_{\mu\nu}=0, \qquad A^{(a)}_{\mu} =0,\qquad M=I_{28}\, .}}
Since the dilaton, the antisymmetric tensor, all 28 gauge fields and
the matrix valued scalar field is trivial, it reduces the equations of motion
of the effective field theory to Einstein equations. Thus the Kerr metric is
a solution to the equations of motion of the effective field theory.
Since we are considering the static solutions, the action is 
three dimensional. It is known that the action has T and S duality 
symmetries that are given by $O(6, 22)$ and $SL(2,Z)$ 
symmetry groups respectively. In fact, these
two groups combine together to a $O(8, 24)$ symmetry group for the 
effective three dimensional action \refs{\ashoke}.
The matrix valued scalar field in this case is given by a $32\times 32$
matrix.
The off-diagonal component of the metric can be treated as gauge fields in
lower dimensions. By dualising gauge fields to scalar fields in three
dimensions we can write a $32\times 32$ matrix
\eqn\dual{\eqalign{
{\cal M} &=\pmatrix{\bar M-e^{2\bar\Phi}\psi\psi^T&e^{2\bar\Phi}\psi&\bar M
\bar L\psi-{1\over 2}e^{2\bar\Phi}\psi(\psi^T\bar L\psi)\cr
&&\cr
e^{2\bar\Phi}\psi^T&-e^{2\bar\Phi}&{1\over 2}e^{2\bar\Phi}\psi^T\bar L\psi\cr
&&\cr
\psi^T\bar L\bar M-{1\over 2}e^{2\bar\Phi}\psi^T(\psi^T\bar L\psi)&
{1\over 2}e^{2\bar\Phi}\psi^T\bar L\psi& -e^{-2\bar\Phi}+\psi^T\bar L\bar M
\bar L\psi\cr &&-{1\over 4}e^{2\bar\Phi}(\psi^T\bar L\psi)^2}\cr
&\cr
{\cal L} &=\pmatrix{\bar L&0&0\cr0&0&1\cr0&1&0}\qquad
{\bar L} =\pmatrix{-I_{22}&0&0&0\cr0&I_6&0&0\cr0&0&0&1\cr0&0&1&0}\qquad
\bar g_{ij} = e^{-2\bar\Phi}\bar G_{ij}.}}
In writing the matrix ${\cal M}$ we used the following duality relations
\eqn\tdual{\eqalign{\bar A^{(a)}_i &=A^{(a)}_i-(G_{tt})^{-1}G_{ti}A^{(a)}_t,
\quad 1\leq a\leq 28,\,\, 1\leq i\leq 3,\cr
\bar A^{(29)}_i &= {1\over 2}(G_{tt})^{-1}G_{ti},\cr
\bar A^{(30)}_i &={1\over 2}B_{ti}+A^{(a)}_tL_{ab}\bar A^{(b)}_i,\cr
\bar G_{ij} &= G_{ij}-(G_{tt})^{-1}G_{ti}G_{tj},\cr
\bar B_{ij} &=
B_{ij}+(G_{tt})^{-1}(G_{ti}A^{(a)}_j-G_{tj}A^{(a)}_i)L_{ab}A^{(b)}_t\cr
&+{1\over 2}(G_{tt})^{-1}(B_{ti}G_{tj}-B_{tj}G_{ti}),\cr
\bar\Phi &=\Phi-{1\over 2}\ln(-G_{tt})}}
\eqn\mbar{\bar M = \pmatrix{M+4(G_{tt})^{-1}A_tA_t^T&-2(G_{tt})^{-1}A_t&
2MLA_t\cr &&+4(G_{tt})^{-1}A_t(A_t^TLA_t)\cr &&\cr-2(G_{tt})^{-1}A_t^T&
(G_{tt})^{-1}&-2(G_{tt})^{-1}A_t^TLA_t\cr &&\cr2A_t^TLM&
-2(G_{tt})^{-1}A_t^TLA_t&G_{tt}+4A_t^TLMLA_t\cr
+4(G_{tt})^{-1}A_t^T(A_t^TLA_t)&&+4(G_{tt})^{-1}(A_t^TLA_t)^2}}
and
\eqn\lbar{\bar L =\pmatrix{L&0&0\cr 0&0&1\cr 0&1&0}}
where $\psi$ is obtained by dualising the gauge fields.
The duality
relation between the gauge fields and $\psi$ is
\eqn\gpsi{\sqrt{\det\bar G}\, e^{-\bar\Phi}(\bar M\bar L)_{\bar a\bar b}
\bar G^{ii'}\bar G^{jj'}\bar F^{(\bar b)}_{i'j'}={1\over 2}\epsilon^{ijk}
\partial_k\psi^{\bar a},}
and the bianchi identity for the field strength can now be written in terms of
$\psi$ as
\eqn\bian{\bar D^i(e^{\bar\Phi}(\bar M\bar L)_{\bar a\bar b}\partial_i
\psi^{\bar b})=0.}
In the above ${\bar a}$ takes values from 1 to 30.
The corresponding three dimensional action from which these equations can be
derived is given by
\eqn\tdact{S_3 = \int d^3x \sqrt{\det\bar g}\,[R_{\bar g}+{1\over 8}\bar
g^{ij}Tr(\partial_i{\cal ML}\partial_j{\cal ML})].}
This action is invariant under the O(8,24) transformation
\eqn\oetf{{\cal M}\rightarrow \bar\Omega{\cal M}\bar\Omega^T,\quad \bar g_{ij}
\rightarrow \bar g_{ij},}
where $\bar\Omega$ is a $32\times 32$ matrix which leaves ${\cal L}$
invariant.

Transformation that diagonalises ${\cal L}$ is,
\eqn\umat{
U =\pmatrix{I_{28}&0&0&0&0\cr 0&{1\over\sqrt{2}}&{1\over\sqrt{2}}&0&0\cr
0&{1\over\sqrt{2}}&{-1\over\sqrt{2}}&0&0\cr 0&0&0&{1\over\sqrt{2}}&
{1\over\sqrt{2}}\cr 0&0&0&{1\over\sqrt{2}}&{-1\over\sqrt{2}}}.}
In the next section we will work in the frame where ${\cal L}$ is diagonal.
This choice, though, is just a matter of convenience.

\newsec{Dyonic Black Hole}

In this section we will give the explicit ansatz of $\Omega$ the matrices
used for duality rotation of ${\cal M}$. We will use the strategy outlined
in the previous section to get the matrix valued scalar field ${\cal M}$
starting from the Kerr solution. After implementing the transformations we
will extract the expressions for various charges characterising the
transformed solution.

Using the duality transformations given in the last section we find that the
Kerr solution can be written as,
\eqn\kerpsi{\eqalign{
\bar g_{ij}dx^idx^j &=(\rho^2+a^2\cos^2{\theta}-2m\rho)[{1\over{\rho^2+a^2
-2m\rho}}d\rho^2+d\theta^2\cr
&+{{\rho^2+a^2-2m\rho}\over{\rho^2+a^2\cos^2{\theta}-
2m\rho}}\sin^2{\theta}d\phi^2]\cr
\psi^a &= \delta_{a,30}{2ma\cos{\theta}\over{\rho^2+a^2\cos^2{\theta}}},}}
and
\eqn\calm{\eqalign{
{\cal M} &=\pmatrix{I_{28}&0&0&0&0\cr 0&-f^{-1}&0&0&-g\cr
0&0&-f-fg^2&g&0\cr 0&0&g&-f^{-1}&0\cr 0&-g&0&0&-f-fg^2}\cr
\tilde{\cal M} &= U{\cal M}U^T = \pmatrix{I_{28}&0&0&0&0\cr 0&A-1&
B&0&g\cr0&B&A-1&-g&0\cr 0&0&-g&A-1&B\cr 0&g&0&B&A-1},}}
where $A = 1-f^{-1}/2-f(1+g^2)/2$, $B = f(1+g^2)/2-f^{-1}/2$ and
\eqn\fg{f = {\rho^2+a^2\cos^2{\theta}-2m\rho\over{\rho^2+a^2\cos^2{\theta}}},
\qquad g = {2ma\cos{\theta}\over{\rho^2+a^2\cos^2{\theta}-2m\rho}}}
\eqn\ab{A={-2m^2\over{\rho^2+a^2\cos^2{\theta}-2m\rho}},\qquad
B={2m(m-\rho)\over{\rho^2+a^2\cos^2{\theta}-2m\rho}}.}
At this point let us also note down the asymptotic behaviour of $A$, $B$,
and $g$ as $\rho\rightarrow\infty$
\eqn\asym{A\sim -{2m^2\over\rho^2}+...\quad B\sim -{2m\over\rho}+...\quad
g\sim {2ma\cos\theta\over\rho^2}+...}
These expressions will be useful for determining the mass, angular momentum
and the charges of the black hole. Matrices used for the duality rotations are
\eqn\mdrot{\eqalign{
\Omega_1 &= \pmatrix{I_{20}&0&0&0&0&0&0&0&0&0\cr 0&O_1&O_2&0&0&0&O_3&0&O_4
&0\cr 0&O_5&O_6&0&0&0&O_7&0&0&0\cr 0&0&0&I_4&0&0&0&0&0&0\cr 0&0&0&0&P_1&
P_2&0&P_3&0&P_4\cr 0&0&0&0&P_5&P_6&0&P_7&0&0\cr 0&O_8&P_8&0&0&0&Q_1&0&Q_2&0\cr
0&0&0&0&0&Q_3&0&Q_4&0&0\cr 0&O_9&P_9&0&0&0&Q_5&0&Q_6&0\cr
0&0&0&0&Q_7&Q_8&0&Q_9&0&Q_{10}}\cr
&\cr
\Omega_2 &= \pmatrix{R_{22}&0&0\cr 0&R_6&0\cr 0&0&I_4},
}}
where,
\eqn\opq{\eqalign{
O_1&=\cb\cT,\, O_2=\ca\cb\sT,\,O_3=\sa\cb\sT\cr
O_4 &=\sb,\, O_5=-\sT,\,O_6 =\ca\cT,\, O_7=\sa\cT\cr
O_8 &=\sb\su\cT,\, O_9 = \sb\cu\cT}}
\eqn\opqc{\eqalign{P_1 &=\cd\cR,\, P_2 = \cg\cd\sR,\,P_3 = \sg\cd\sR\cr
P_4 &=\sd,\, P_5=-\sR,\,P_6=\cg\cR,\,P_7=\sg\cR\cr
P_8 &= \ca\sb\su\sT+\sa\cu\cr
P_9 &= \ca\sb\cu\sT-\sa\su}}
\eqn\opqt{\eqalign{Q_1 &= \sa\sb\su\sT+\ca\cu,\,Q_2 = \cb\su\cr
Q_3 &=\sg,\, Q_4 =\cg,\,Q_5 = \sa\sb\cu\sT-\ca\su\cr
Q_6 &=\cb\cu,\, Q_7 =\sd\cR,\,Q_8 =\cg\sd\sR\cr
Q_9 &= \sg\sd\sR,\, Q_{10} =\cd}}
and $R_N$ denotes any $N$-dimensional rotation matrix.
Some comments regarding the choice of these matrices are in order. The choice
of the matrix $\Omega_1$ is made in a following way. It was noticed earlier
\refs{\sen} that the most general transformation that preserves asymptotic
form of all the field configurations and gives inequivalent solutions belongs
to the coset
\eqn\coset{(O(22,2)\times O(6,2))/(O(22)\times O(6)\times SO(2)).}
We choose to perform this transformation in five steps. The matrix $\Omega_1$
contains four of them, two hyperbolic rotations corresponding to
$O(22,2)/O(22)$ and similarly for $O(6,2)/O(6)$. These transformations are
chosen in such a way that the electric and the magnetic charge vectors lie in
a plane with non-zero dot and cross product. We then embed this plane in
$22$ and $6$ directions by a general rotation performed by $\Omega_2$.

After doing the duality rotation of ${\cal M}$ with respect to both
$\Omega$s we can extract the expressions for various fields in the new
solution. Most of these expressions are very lenghty and cumbersome
\foot{Details will be published elsewhere\refs{\jmp}, however, the
spherically symmetric case is discussed in detail in section (5).}
, but to determine exact
expressions for the mass, the angular momentum and the charges it suffices
to look at the asymptotic form of these fields. Substituting the asymptotic
form of $A$, $B$ and $g$ in the expressions of $G_{tt}$, $\Phi$, $\psi$ and
$A_t$ we find that the black hole solution obtained using \mdrot\ has

\noindent Mass:\foot{notations are given in the appendix}
\eqn\mass{{\rm Mass} = {m\over 2}\Delta^{1/2}}
Angular Momentum:
\eqn\angmom{\eqalign{J &= {ma\over 2\Delta^{1/2}}(\ca\cb(s_r^2)
+\cg\cd(s_t^2)\cr
&+(\ca\cg+\cb\cd)(\ca\cd+\cb\cg))}}
Electric Charge:($1\leq n\leq 22$ and $1\leq p\leq 6$)
\eqn\elchg{\eqalign{
Q^{(n)}_{el}&={m\over\sqrt{2}}R_{22}\pmatrix{0_{20}\cr \cr O_3Q_4\cr \cr
O_7Q_4}\cr
&\cr
Q^{(p)}_{el}&={m\over\sqrt{2}}R_6\pmatrix{0_4\cr \cr P_3Q_1+P_4Q_2\cr \cr
P_7Q_1}}}
Magnetic charge:($1\leq n\leq 22$ and $1\leq p\leq 6$)
\eqn\mgchg{\eqalign{
Q^{(n)}_{mag}&={m\over\sqrt{2}}R_{22}\pmatrix{0_{20}\cr \cr O_3Q_9+O_4Q_{10}
\cr \cr O_7Q_9}\cr
&\cr
Q^{(p)}_{mag}&={-m\over\sqrt{2}}R_{6}\pmatrix{0_{4}\cr \cr P_3Q_5+P_4Q_6\cr
\cr P_7Q_5}}}
Electric dipole moment:($1\leq n\leq 22$ and $1\leq p\leq 6$)
\eqn\edm{\eqalign{\mu_{el}^{(n)} &={ma\over\sqrt{2}}R_{22}\pmatrix{0_{20}\cr
\cr O_4Q_4\cr \cr 0}\cr
\mu_{el}^{(p)} &= {ma\over\sqrt{2}}R_6\pmatrix{0_4\cr \cr P_3Q_2-P_4Q_1\cr
\cr P_7Q_2}}}
Magnetic dipole moment:($1\leq n\leq 22$ and $1\leq p\leq 6$)
\eqn\mdm{\eqalign{\mu_{mag}^{(n)} &={ma\over\sqrt{2}}R_{22}\pmatrix{0_{20}\cr
\cr O_3Q_{10}-O_4Q_9\cr \cr O_7Q_{10}}\cr
\mu_{mag}^{(p)} &={ma\over\sqrt{2}}R_6\pmatrix{0_4\cr \cr P_3Q_6-P_4Q_5\cr
\cr P_7Q_6}}}
where $R_{22}$ and $R_6$ are 22 and 6 dimensional rotation matrices.
It is known that when we use the equation \mdrot\ to get new solutions with the
magnetic charge it also generates the NUT charge, which corresponds to the
singularity in the $G_{t\phi}$ component of the metric. Since we are
interested in the non-singular metrics, the NUT charge of the new black hole
should be zero. The condition for vanishing of the NUT charge is
\eqn\nonut{
\tan{u}={\sa\sb\cg\sT-\ca\sg\sd\sR\over \sa\sb\sg\sd\sT\sR+\cb\cd+\ca\cg}.}
All the charges as well as the moments carried by the black hole are
subject to this constraint. We will impliment this constraint in the rest
of the paper.
Since the $G_{\rho\rho}$ part of the metric in the heterotic string frame is
unaltered the space-time structure remains same as the original Kerr solution,
i.e. this solution in general has two horizons which are
located at
\eqn\hor{\rho=m\pm\sqrt{m^2-a^2}.}
The case $a > m$ corresponds to the naked singularity whereas $a\rightarrow
m$ is the extremal limit. The curvature singularity is time-like and is
concentrated on a ring at $\phi = 0$ and $\theta = \pi/2$.

\newsec{Various Limits}

We can take various limits of the black hole solution described in the
previous section by changing parameters. In particular, we will see that
in certain limit this solution can be reduced either to pure electrically
charged black hole or pure magnetically charged black hole solution. We will
examine the extremal limits in this section. We will show that in certain
limits this solution saturates the Bogomolnyi bound. 
We should point out at this stage that saturation of Bogomolnyi
bound is a necessary condition for the solution to be supersymmetric,
but it is yet not so clear if it is 
sufficient~\refs{\rk}. One would thus need to figure out 
the killing spinor in 
order confirm the solution to be supersymmetric.
However, in most of the cases in the past, solutions saturating
Bonomolnyi bound turned out to preserve partial supersymmetry.

Case I: $\alpha\rightarrow\infty,~m\rightarrow 0,~a
\rightarrow 0$ such that $m\ca\equiv
m_0$ finite

\eqn\mca{\eqalign{{\rm Mass} &= {m_0\over 2}[N_1(\cg+\sb\sg\sd\sT\sR)\cr
&+ N_2(\sb\cg\sT-\sg\sd\sR)]\cr
&\cr
Q^{(n)}_{el} &={m_0\over\sqrt{2}}R_{22}\pmatrix{0_{20}\cr \cr\cb\cg\sT\cr
\cr\cg\cT},
\cr
Q^{(n)}_{mag}&={m_0\over\sqrt{2}}R_{22}\pmatrix{0_{20}\cr \cr\cb\sg\sd\sT
\sR\cr \cr\sg\sd\cT\sR},}}
\eqn\mcaa{\eqalign{Q^{(p)}_{el} &={m_0\over\sqrt{2}}R_{6}\pmatrix{0_{4}\cr \cr
N_1\sg\cd\sR\cr +N_2\sb\sg\cd\sT\sR\cr \cr N_1\sg\cR\cr +N_2\sb\sg\sT\cR}\cr
Q^{(p)}_{mag}&={-m_0\over\sqrt{2}}R_{6}\pmatrix{0_{4}\cr
\cr N_1\sb\sg\cd\sT\sR\cr-N_2\sg\cd\sR\cr \cr N_1\sb\sg\sT\cR\cr -N_2\sg\cR}.}}
where in the above and in the rest of the paper we have defined
$N_1 = ((s_t)(s_r)+\cb\cd +\ca\cg)\Delta^{-1/2}$ and
$N_2 = ((s_t)\cg - (s_r)\ca)\Delta^{-1/2}$. 
All other quatities like electric and magnetic moments, angular momentum
vanish since $a$ is taken to zero.
In this limit, the relation between the mass and the charges is given by
\eqn\mchrga{({\rm Mass})^2 = {1\over 2}[(Q^{(n)}_{el})^2+(Q^{(n)}_{mag})^2],}
and $[(Q^{(p)}_{el})^2+(Q^{(p)}_{mag})^2] < [(Q^{(n)}_{el})^2+
(Q^{(n)}_{mag})^2]$, therefore  
case I does not lead to a black hole preserving half supersymmetry. 
Nevertheless, it can, in principle, preserve 1/4 supersymmetry. The mass
formula for the latter case\refs{\cybps} in our context is given by
\eqn\quart{M^2_{BPS} = Q^{(p)T}_{el} Q^{(p)}_{el}
+Q^{(p)T}_{mag} Q^{(p)}_{mag}+2[Q^{(p)T}_{el} Q^{(p)}_{el}
Q^{(p)T}_{mag} Q^{(p)}_{mag}-(Q^{(p)T}_{el} Q^{(p)}_{mag})^2]^{1\over 2}.}
We find that this configuration does not satisfy the 1/4 supersymmetry 
constraint as well. Thus, we conclude that this black hole breaks all the
supersymmetry.

case II:$\beta\rightarrow\infty,~m\rightarrow 0,~a\rightarrow 0$ 
such that $m\cb\equiv
m_0$ finite

\eqn\mcb{\eqalign{{\rm Mass} &= {m_0\over 2}[N_1(\cd+\sa\sg\sd\sT\sR)\cr
&+N_2\sa\cg\sT]\cr
&\cr
Q^{(n)}_{el} &={m_0\over\sqrt{2}}R_{22}\pmatrix{0_{20}\cr \cr\sa\cg\sT\cr \cr
0},\cr
Q^{(n)}_{mag}&={m_0\over\sqrt{2}}R_{22}\pmatrix{0_{20}\cr \cr\cd+\sa\sg\sd\sT
\sR\cr \cr0},}}
\eqn\mcba{\eqalign{Q^{(p)}_{el} &={m_0 N_2\over\sqrt{2}}R_{6}\pmatrix{0_{4}\cr
 \cr (\sd+\sa\sg\cd\sT\sR)\cr \cr\sa\sg\sT\cR}\cr
&\cr
Q^{(p)}_{mag}&={-m_0 N_1\over\sqrt{2}}R_{6}\pmatrix{0_{4}\cr \cr (\sa\sg\cd\sT
\sR+\sd)\cr \cr \sa\sg\sT\cR}.}}
All other physical quantities vanish. 
In this limit, the relation between the mass and the charges is again given by
\eqn\mchrgb{({\rm Mass})^2 = {1\over 2}[(Q^{(n)}_{el})^2+(Q^{(n)}_{mag})^2],}
and $[(Q^{(p)}_{el})^2+(Q^{(p)}_{mag})^2] < [(Q^{(n)}_{el})^2+
(Q^{(n)}_{mag})^2]$, therefore this limit too does not correspond to a
black hole preserving half supersymmetry. It is easy to check that it also 
fails to satisfy the general BPS mass formula \quart\ and hence breaks all
supersymmetries.

case III:$\gamma\rightarrow\infty,~m\rightarrow 0,~a\rightarrow 0$ 
such that $m\cg\equiv
m_0$ finite

The nonzero physical quantities are:
\eqn\mcg{\eqalign{{\rm Mass} &= {m_0\over 2}[N_1(\ca+\sa\sb\sd\sT\sR)\cr
&+N_2(\sa\sb\sT-\ca\sd\sR)]\cr
&\cr
Q^{(n)}_{el} &={m_0\over\sqrt{2}}R_{22}\pmatrix{0_{20}\cr\sa\cb\sT\cr \sa\cT},
\cr Q^{(n)}_{mag}&={m_0\over\sqrt{2}}R_{22}\pmatrix{0_{20}\cr\sa\cb\sd\sT\sR
\cr \sa\sd\sR\cT}}}
\eqn\mcga{\eqalign{Q^{(p)}_{el} &={m_0\over\sqrt{2}}R_{6}\pmatrix{0_{4}\cr
N_1\ca\cd\sR\cr +N_2\sa\sb\cd\sT\sR\cr \cr N_1\ca\cR\cr +N_2\sa\sb\sT\cR}\cr
&\cr
Q^{(p)}_{mag}&={-m_0\over\sqrt{2}}R_{6}\pmatrix{0_{4}\cr N_1\sa\sb\cd\sT\sR\cr
-N_2\ca\cd\sR\cr \cr N_1\sa\sb\sT\cR\cr -N_2\ca\cR}.}}
In this limit, the relation between the mass and the charges is
\eqn\mchrgg{({\rm Mass})^2 = {1\over 2}[(Q^{(p)}_{el})^2+(Q^{(p)}_{mag})^2],}
therefore in the extremal limit this solution corresponds to a BPS saturated
black hole carrying both electric and magnetic charge. This is an extremal
dyonic black hole which is specified by 57 parameters, 28 dimensional
electric charge vector, 28 dimensional magnetic charge vectors and the mass
of the black hole. 

Let us study this limit in more details. If we take the limit $\sR\rightarrow
0$, this solution reduces to a pure electrically charged solution, i.e.,
$({\rm Mass})^2 = (Q^{(p)}_{el})^2/2$. If we also take $\sT\rightarrow 0$ we
recover the supersymmetric solution(case II) of \refs{\sen}.

case IV:$\delta\rightarrow\infty,~m\rightarrow 0,~a\rightarrow 0$ 
such that $m\cd\equiv
m_0$ finite

The non-zero physical quantities are:
\eqn\mcd{\eqalign{{\rm Mass} &= {m_0\over 2}[N_1(\cb+\sa\sb\sg\sT\sR)\cr
&-N_2\ca\sg\sR]\cr
&\cr
Q^{(n)}_{el} &=0,
\quad Q^{(n)}_{mag}={m_0\over\sqrt{2}}R_{22}\pmatrix{0_{20}\cr \cr\sa\cb\sg
\sT\sR+\sb\cr \cr\sa\sg\sR\cT}}}
\eqn\mcda{\eqalign{Q^{(p)}_{el} &=0,
Q^{(p)}_{mag}={-m_0\over\sqrt{2}}R_{6}\pmatrix{0_{4}\cr \cr N_1(\sa\sb\sg
\sT\sR+\cb)\cr-N_2\ca\sg\sR\cr \cr 0}.}}
In this limit, the relation between the mass and the charges is
\eqn\mchrgd{({\rm Mass})^2 = {1\over 2}(Q^{(p)}_{mag})^2,}
which satisfies the Bogomolnyi bound for mass and
therefore in the extremal limit this solution may lead to a supersymmetric
black hole carrying only magnetic charge. This is an extremal magnetically
charged black hole which is specified by 28 dimensional magnetic charge
vectors and the mass of the black hole.

case V:$\alpha=\gamma\rightarrow\infty,~m\rightarrow 0,~a\rightarrow 0$ 
such that
$m\cta\equiv m_0$ finite

The non-zero physical quantities in this limit are:
\eqn\mcag{\eqalign{{\rm Mass} &= {m_0\over 2}[N_1(1+\sb\sd\sT\sR)\cr
&+N_2(\sb\sT-\sd\sR)]\cr
&\cr
Q^{(n)}_{el} &={m_0\over\sqrt{2}}R_{22}\pmatrix{0_{20}\cr\cb\sT\cr \cT},
\, Q^{(n)}_{mag}={m_0\over\sqrt{2}}R_{22}\pmatrix{0_{20}\cr\cb\sd\sT\sR
\cr \sd\sR\cT}}}
\eqn\mcaga{\eqalign{Q^{(p)}_{el} &={m_0\over\sqrt{2}}R_{6}\pmatrix{0_{4}\cr
N_1\cd\sR+N_2\sb\cd\sT\sR\cr N_1\cR+N_2\sb\sT\cR}\cr
&\cr
Q^{(p)}_{mag}&={-m_0\over\sqrt{2}}R_{6}\pmatrix{0_{4}\cr N_1\sb\cd\sT\sR-
N_2\cd\sR\cr N_1\sb\sT\cR-N_2\cR}.}}
In this limit, the relation between the mass and the charges is
\eqn\mchag{({\rm Mass})^2 = {1\over 2}[(Q^{(p)}_{el})^2+(Q^{(p)}_{mag})^2]
= {1\over 2}[(Q^{(n)}_{el})^2+(Q^{(n)}_{mag})^2],}
therefore again this is BPS saturated and may lead to a 
supersymmetric dyonic black hole carrying both electric and magnetic charge. 
This is an extremal dyonic black hole which is specified by 57 parameters, 
28 dimensional electric charge vector, 28 dimensional magnetic charge vectors 
and the mass of the black hole.

case VI:$\beta=\delta\rightarrow\infty,~m\rightarrow 0,~a\rightarrow 0$ 
such that
$m\ctb\equiv m_0$ finite

\eqn\mcbd{\eqalign{{\rm Mass} &= {m_0 N_1\over 2}(1+\sa\sg\sT\sR)\cr
&\cr
Q^{(n)}_{el} &=0,
\qquad Q^{(n)}_{mag}={m_0\over\sqrt{2}}R_{22}\pmatrix{0_{20}\cr 1+\sa\sg\sT\sR
\cr 0}}}
\eqn\mcbda{\eqalign{Q^{(p)}_{el} &=0 \qquad
Q^{(p)}_{mag}={-m_0\over\sqrt{2}}R_{6}\pmatrix{0_{4}\cr 1+\sa\sg\sT\sR\cr
0}}.}
All the other physical quantities such as electric and magnetic 
moments and angular momentum vanish.
In this limit, the relation between the mass and the charges is
\eqn\maschrgbd{({\rm Mass})^2 = {1\over 2}(Q^{(p)}_{mag})^2={1\over 2}
(Q^{(n)}_{mag})^2.}
Here also mass saturates the Bogomolnyi bound and hence this solution is a 
potential candidate for a supersymmetric
black hole carrying only magnetic charge. This is an extremal magnetically
charged black hole which is specified by 28 dimensional magnetic charge
vectors and the mass of the black hole.

\newsec{Non-rotating Dyonic Black Hole}

In this section we give the explicit solutions for the non-rotating dyonic
black holes by restricting the Kerr metric to the case $a=0$. After
carrying out the duality transformations, as discussed earlier, we find
the metric to be
\eqn\nrmetric{\eqalign{G_{\mu\nu} dx^\mu dx^\nu &= G_{tt} dt^2 +
({\rho^2 -2m\rho(1-b_{\Delta})+m^2(a_{\Delta}-2b_{\Delta})\over
{\rho^2-2m\rho}})(d\rho^2\cr
&+(\rho^2- 2m\rho)(d\theta^2 + \sin^2\theta d\phi^2))}}
where (notations used in this section are explained in the appendix)
\eqn\gtt{G_{tt} = -{(\rho^2-2m\rho)(\rho^2-2m\rho(1-\bd)+m^2(\ad-2\bd))\over
\rho^4+2m\rho^3d_3+m^2\rho^2d_2+2m^3\rho d_1+m^4d_0}}
and the dilaton is given by
\eqn\dilt{\Phi = {1\over 2}\ln({\rho^4-4m\rho^3 n_3+2m^2\rho^2 n_2
-4m^3\rho n_1+m^4n_0\over
\rho^4+2m\rho^3d_3+m^2\rho^2d_2+2m^3\rho d_1+m^4d_0}) =
{1\over 2}\ln{\Phi_n\over\Phi_d}.}
The time components of the gauge fields are 
\eqn\gaut{\eqalign{A_t^{(n)}&={R_{22}\over\sqrt{2}\Phi_d}\pmatrix{0_{20}\cr 
G_n M_1-G_p\psi_1\cr G_n M_2-G_p\psi_2}\cr
A_t^{(p)}&={R_6\over\sqrt{2}\Phi_d}\pmatrix{0_4\cr G_n M_3-G_p\psi_3\cr
G_n M_4-G_p\psi_4}}}
where,
\eqn\gs{\eqalign{G_n &= \r^2-2m\r(1-\bd)+m^2(\ad-2\bd)\cr
G_p &= m^2(a_1-b_1+\cg(s_r))+m\r b_1}}
with
\eqn\mforat{\eqalign{M_1 &= -{m^2\over\Delta^{1/2}}\{(s_t)\coth\beta(\ca\cb
\cd+\cta\cg+(s_t^2)\cg)\cr
&+\cb\sb(\cg(s_t)-\ca(s_r))\}+m(m-\r)(s_t)\cg\coth\beta\cr
M_2 &= \cT\sa[-{m^2\over\Delta^{1/2}}(\ca\cb\cd+\cta\cg\cr
&+(s_t^2)\cg)+m(m-\r)\cg]\cr
M_3 &= -m^2(s_r)\cg\coth\delta+{m(m-\r)\over\Delta^{1/2}}
\{\coth\delta(s_r)(\ca\cb\cd\cr 
&+\cta\cg+\cg(s_t^2)+\cb\sd(\cg(s_t)-\ca(s_r)))\}\cr
M_4 &= \cR\sg[-m^2\cg+{m(m-\r)\over\Delta^{1/2}}(\ca\cb\cd\cr
&+\cta\cg+\cg(s_t^2))],}}
and
\eqn\psifort{\eqalign{\psi_1 = &-m^2(b_4\cb\sb+b_5\coth\beta)
+m(m-\r)(\sb\cd\cr
&+(s_ts_r)\coth\beta)\cr
\psi_2 &= \cT\sa[-{m^2b_1\over 2\cg}+m(m-\r)(s_r)]\cr
\psi_3 &= -m^2(\cd\sd(1+\stg\stR))+m(m-\r)(b_2\tanh\delta\cr
&+b_3\coth\delta)\cr
\psi_4 &= \cR[-m^2\sg(s_r)+{m(m-\r)b_1\over 2}\tanh\gamma].}}
Similarly the spatial components of the gauge fields are
\eqn\gauphi{\eqalign{A_{\phi}^{(n)}&={-R_{22}\cos\theta\over 2(\r^2-2m\r)G_n}
\pmatrix{0_{20}\cr -\bar M_1\Psi_1-\bar M_2\Psi_2+\bar
M_3\Psi_3\cr +\bar M_4\Psi_4+\bar M_5\Psi_5
+ \bar M_6\Psi_6\cr \cr 
-\bar M_2\Psi_1-\bar M_5\Psi_2+\bar
M_8\Psi_3\cr +\bar M_9\Psi_4+\bar M_{10}\Psi_5
+ \bar M_{11}\Psi_6}\cr
A_{\phi}^{(p)}&={-R_6\cos\theta\over 2(\r^2-2m\r)G_n}\pmatrix{0_4\cr 
-\bar M_3\Psi_1-\bar M_8\Psi_2+\bar
M_{12}\Psi_3\cr +\bar M_{13}\Psi_4+\bar M_{14}\Psi_5
+ \bar M_{15}\Psi_6\cr \cr 
-\bar M_4\Psi_1-\bar M_9\Psi_2+\bar
M_{13}\Psi_3\cr +\bar M_{16}\Psi_4+\bar M_{17}\Psi_5
+ \bar M_{18}\Psi_6}}}
where $\bar M$'s and $\Psi$'s are defined in the appendix.
The moduli fields are given by
\eqn\moduli{M = I_{28} +\pmatrix{R_{22}PR^T_{22}& R_{22}QR^T_6\cr
(R_{22}QR^T_6)^T& R_6RR^T_6}}
where,
\eqn\modpqr{\eqalign{P &=\pmatrix{0_{20\times 20}& 0& 0\cr 0& p_1& p_2\cr
0& p_2& p_3}\cr
Q &=\pmatrix{0_{4\times 20}& 0& 0\cr 0& q_1& q_2\cr 0& q_3& q_4}\cr
R &=\pmatrix{0_{4\times 4}& 0& 0\cr 0& r_1& r_2\cr 0& r_2& r_3}.}}
The variables $p_i$, $q_i$ and $r_i$ can be written in terms of $\bar M$,
$G_n$, $\Psi$ and $\Phi$ and are given in the appendix. The axion, a 
pseudoscalar field $\chi$, is related to the three form field strength
$H^{\mu\nu\lambda}=-\exp{(2\Phi)}(-G)^{-1/2}\epsilon^{\mu\nu\lambda\sigma}
\partial_{\sigma}\chi$ and is given by
\eqn\axi{\eqalign{\chi &= {\sin^2\theta\over 4\sqrt{2}}\int {d\rho\over
(\rho^2-2m\rho)^2\sqrt{\Phi_d}}\cr
&[(G_n M_1-G_p\psi_1)(-\bar M_1\Psi_1-\bar M_2\Psi_2+\bar M_3\Psi_3+
\bar M_4\Psi_4+\bar M_5\Psi_5+\bar M_6\Psi_6)\cr
&\!\!\!+(G_n M_2-G_p\psi_2)(-\bar M_2\Psi_1-\bar M_5\Psi_2+\bar M_8\Psi_3+
\bar M_9\Psi_4+\bar M_{10}\Psi_5+\bar M_{11}\Psi_6)\cr
&\!\!\!\!-(G_nM_3-G_p\psi_3)(-\bar M_3\Psi_1-\bar M_8\Psi_2+\bar M_{12}\Psi_3+
\bar M_{13}\Psi_4+\bar M_{14}\Psi_5+\bar M_{15}\Psi_6)\cr
&\!\!\!\!\!-(G_nM_4-G_p\psi_4)(-\bar M_4\Psi_1-\bar M_9\Psi_2+\bar M_{13}
\Psi_3+
\bar M_{16}\Psi_4+\bar M_{17}\Psi_5+\bar M_{18}\Psi_6)]}}

The electric and magnetic charges are identical to those mentioned in the
section (3) however, both electric and magnetic dipole moments as well as 
the angular
momentum vanish. The Einstein metric is given by
\eqn\emet{\eqalign{g_{\mu\nu}dx^{\mu}dx^{\nu} &= e^{-\Phi}G_{\mu\nu}dx^{\mu}
dx^{\nu} = ({\Phi_d\over\Phi_n})^{1/2}(G_{tt}dt^2+\cr
&{G_n\over \rho^2-2m\rho}(d\rho^2+
(\rho^2-2m\rho)(d\theta^2+\sin^2\theta d\phi^2)))}.}
Using this metric it is easy to see that the event horizon is located at
$\rho=2m$ and the area of the event horizon is given by
\eqn\area{\eqalign{A &= \int d\theta d\phi ({\Phi_d\over\Phi_n})^{1/2}
G_n\sin\theta|_{\rho=2m}\cr
&=4\pi m^2(4\ca\cb\cg\cd-2c_4-c_3)^{1/2}.}}
The surface gravity is of the black hole is given by,
\eqn\sgr{\kappa =\lim_{\rho\rightarrow 2m}\sqrt{g^{\rho\rho}}
\partial_{\rho}\sqrt{-g_{tt}} = {1\over m(4\ca\cb\cg\cd-2c_4-c_3)^{1/2}}.}
In the limit   $m\rightarrow 0,$ 
event horizon touches the singularity. Various limits of the boost
parameters can be considered for this solution, which is similar in spirit
to previous section. All the limiting cases of section (3) are trivially 
applicable to this
solution. This solution therefore represents a non-rotating black hole
solution carrying 28 electric and 28 magnetic charges. In the limit
$\beta$, $\delta$, $R$ and $T\rightarrow 0$ our results reduces to the
results those of \refs{\sen} in the $a\rightarrow 0$ limit.

\newsec{Conclusion}

In this paper, we have presented a general black hole solution to the
heterotic string theory compactified on a six dimensional torus.  We also
studied the solution in various limits and we find that in certain cases
the black hole mass saturates the Bogomolnyi bound. 
One of the motivations behind
constructing this solution is to study if black hole can have elementary
particle like behaviour. We studied the non-rotating black hole in detail.
All the limits studied in the earlier sections also apply to this solution
as well. We thus find the most general non-rotating black hole solution in
the heterotic string theory on a torus which carries 28 electric and 28
magnetic charges.

\bigskip
While preparing this manuscript, a recent preprint ~\refs{\cvetic}
appeared which addresses similar issues.
\bigskip
\noindent{\bf Acknowledgements:} We would like to thank A. Sen for several
discussions and R. Kallosh for a fruitful communication.
 SP acknowledges the hospitality of ICTP, Trieste, during which
this work was initiated. Work of SM was partially supported by NSF Grant
No. PHY-9411543
\vfill

\newsec{Appendix}

This appendix contains the shorthand notation used in the paper:

\eqn\notation{\eqalign{(s_t) &= \sa\sb\sT\quad (s_r) = \sg\sd\sR\cr
\Delta &= (s_t^2s_r^2)+2\cb\cd(s_ts_r)+\cta(s_r^2)\cr
&+\ctg(s_t^2)+(\ca\cg+\cb\cd)^2\cr
a_{\Delta} 
&=\Delta^{-1}[(s_t^4s_r^2)+(s_t^2s_r^4)+\cta(s_r^4)+(s_t^2s_r^2)(2\cta+\ctb\cr
&+\ctg+\ctd)+2\cb\cd(s_t^3s_r+s_ts_r^3)\cr
&+(\ctb\ctd+\ctd\ctg)(s_t^2)+(\cfa+\cta\ctd\cr
&+\cta\ctg+\ctb\ctd+2\ca\cb\cg\cd)(s_r^2)\cr
&+2\cb(\cta\cd+\ca\cb\cg+\ctb\cd\cr
&+\ccd)(s_ts_r)+(\cta\ctg+\ctb\ctd)(\ctb+\ctd)\cr
&+2\ca\ccb\cg\cd+2\ca\cb\cg\ccd]\cr
b_{\Delta} &= \Delta^{-1/2}[(s_t^2s_r^2)+2\cb\cd(s_ts_r)\cr
&+\cta(s_r^2)+(\ca\cg+\cb\cd)\cb\cd]\cr
c_1 &= \cta+\ctb+\ctg+\ctd+(s_t^2)+(s_r^2)\cr
&+4\ca\cb\cg\cd\qquad c_2 = \Delta^{1/2}\cr
c_3 &= \Delta^{-1}[-\ctg\ctd(s_t^4)-(2\ctb\ctg\ctd\cr
&+\cta\ctb\ctg+\cfg\ctd\cr
&+2\ca\cb\cg\ccd+2\cta\ctg\ctd)(s_t^2)\cr
&-(2\cta\ctb\ctd+\cfa\ctb+\cta\ctg\ctd\cr
&+2\ca\ccb\cg\cd+2\cta\ctb\ctg)(s_r^2)\cr
&+(4\ca\ctb\cg\ctd-2\cta\ccb\cd\cr
&-2\cb\ctg\ccd)(s_ts_r)-(\cta\ctb+\ctb\ctg\cr
&+\ctg\ctd+\ctd\cta)(\ca\cg+\cb\cd)^2\cr
&-\cta\ctb(s_r^4)-(\cta\ctb+\ctg\ctd)(s_t^2s_r^2)]}}
\eqn\cfor{\eqalign{c_4 &= -\Delta^{-1/2}(\ca\cb+\cg\cd)[\ca\cb(\ctg\cr
&+\ctd+(s_r^2))+\cg\cd(\cta+\ctb+(s_t^2))]\cr
d_0 &= 2c_4-c_3+4\ca\cb\cg\cd\quad d_1 = 2c_2-c_1-c_4\cr
d_2 &= 4+c_1-6c_2\quad d_3 = c_2-2\quad n_0 = (\ad-2\bd)^2\cr
n_1 &= \ad+\bd(2\bd-\ad-2)\quad n_2 = 2-6\bd+2\bd^2+\ad\quad n_3 = 1-\bd
}}
\eqn\ab{\eqalign{a_1 &= {1\over\Delta}[\cg(s_t^4s_r)+\cb\cg\cd(s_t^3)\cr
&+(2\cta\cg+\ca\cb\cd+\ctb\cg)(s_t^2s_r)\cr
&-\ca\ctb(s_ts_r^2)+(\ca\ctb(\ctg+\ctd)\cr
&+\cb\cg\cd(\cta+\ctb))(s_t)\cr
&+\ca(\cta-\ctb)(\ca\cg+\cb\cd)(s_r)]\cr
b_1 &= {2\cg\over\Delta^{1/2}}[(s_t^2s_r)+\cta(s_r)+\cb\cd(s_t)]\cr
b_2 &={1\over\Delta^{1/2}}[\cb\cd(\ca\cg+\cb\cd+(s_ts_r))]\cr
b_3 &= {1\over\Delta^{1/2}}[(s_t^2s_r^2)+\cb\cd(s_ts_r)+\ca(s_r^2)]\cr
b_4 &= {1\over\Delta^{1/2}}[(s_ts_r)+\cb\cd+\ca\cg]\cr
b_5 &= {1\over\Delta^{1/2}}[(s_t^3s_r)+\cb\cd(s_t^2)+\cta(s_ts_r)]}}
\eqn\newms{\eqalign{\bar M_1 &=1+{2\over \r^2-2m\r}[-m^2((s_t)^2
\coth^2\beta+\stb)+{1\over G_n}[m^2(b_4\sb\cr
&\cb+b_5\coth\beta)-m(m-\r)(\sb\cd+(s_ts_r)\coth\beta)]^2]\cr
\bar M_2 &={1\over \r^2-2m\r}[-2m^2\cb\sta\cT\sT\cr
&+{\cT\sa\over G_n}[m^2(b_4\sb\cb
+b_5\coth\beta)-m(m-\r)\cr
&(\sb\cd+(s_ts_r)\coth\beta)][{m^2\over\cg}b_1+2m(m-\r)(s_r)]]}}
\eqn\onemore{\eqalign{\bar M_3 &={1\over 
\r^2-2m\r}[2m(m-\r)(\sb\sd-\sT\sR\sa\cb\cr
&\sg\cd)+{1\over G_n}[m^2(b_4\sb\cb+b_5\coth\beta)-m(m-\r)\cr
&(\sb\cd+(s_ts_r)\coth\beta)][-2m^2\sd\cd(1+\stg\stR)\cr
&+2m(m-\r)(b_2\tanh\delta+b_3\coth\delta)]]\cr
\bar M_4 &={1\over \r^2-2m\r}[-2m(m-\r)
\sa\cb\sg\cR\sT\cr
&+{\cR\over G_n}[m^2(b_4\sb\cb
+b_5\coth\beta)-m(m-\r)(\sb\cd\cr
&+(s_ts_r)\coth\beta)][-2m^2\sg(s_r)+m(m-\r)\tanh\gamma b_1]]\cr
\bar M_5 &={1\over\sqrt{2}(\r^2-2m\r)}[-2m(m-\r)\sa\cb\cg\sT\cr
&-{2m^2\over\sqrt\Delta}\{\sa\ca\cb\sT((s_ts_r)+\cb\cd+\ca\cg)\cr
&+\cb\sb(1+\sta\stT)(\cg(s_t)-\ca(s_r)\}\cr
&+{m^2(a_1-b_1+\cg(s_r))+m\r b_1\over G_n}
[2m^2(b_4\sb\cb+b_5\coth\beta)\cr
&-2m(m-\r)(\sb\cd+(s_ts_r)\coth\beta)]]}}
\eqn\newmsco{\eqalign{
\bar M_6 &={1\over\sqrt{2}(\r^2-2m\r)}[-2m(m-\r)\sa\cb\cg\sT\cr
&-{2m^2\over\sqrt\Delta}\{\sa\ca\cb\sT((s_ts_r)+\cb\cd+\ca\cg)\cr
&+(\cg(s_t)-\ca(s_r)(\cb\sb+(s_t)^2\coth\beta)\}\cr
&+{m^2(a_1-\cg(s_r))\over G_n}
[2m^2(b_4\sb\cb+b_5\coth\beta)\cr
&-2m(m-\r)(\sb\cd+(s_ts_r)\coth\beta)]]\cr
\bar M_7 &= 1+{\ctT\sta\over \r^2-2m\r}(-2m^2+
{1\over 2G_n}(-{m^2\over\cg}b_1+2m(m-\r)(s_r))^2)\cr
\bar M_8 &= {2\cT\sa\over \r^2-2m\r)}[m(m-\r)\sR\sg\cd+{1\over 2G_n}
({m^2b_2\over\cg}\cr
&-2m(m-\r)(s_r))(m^2\sd\cd(1+\stg\stR)\cr
&-m(m-\r)(b_2\tanh\delta+b_3\coth\delta)]}}
\eqn\mscont{\eqalign{\bar M_9 &= {\cR\cT\sa\sg\over \r^2-2m\r}[2m(m-\r)
+{1\over 2G_n}\cr
&({m^2b_1\over \cg}-2m(m-\r)(s_r))(2m^2(s_r)-{m(m-\r)b_1\over\cg})]\cr
\bar M_{10} &= {\sqrt{2}\cT\sa\over \r^2-2m\r}
[m(m-\r)\cg-{m^2\over\sqrt{\Delta}}(\ca\cb\cd\cr
&+\cta\cg+\cg(s_t)^2)+{G_p\over 2G_n}({m^2b_1\over\cg}-2m(m-\r)(s_r))]\cr
\bar M_{11} &= -{\sqrt{2}\cT\sa\over \r^2-2m\r}[m(m-\r)\cg+{m^2\over
\sqrt{\Delta}}(\ca\cb\cd\cr
&+\cta\cg+\cg(s_t)^2)-{1\over 2G_n}[m^2(a_1-\cg(s_r))({m^2b_1\over\cg}\cr
&-2m(m-\r)(s_r))]\cr
\bar M_{12} &= 1-{2m^2\over \r^2-2m\r}[(\std+(s_r)^2\coth^2\delta)
+{1\over G_n}[m\cd\sd(1\cr
&+\stg\stR)-(m-\r)(b_2\tanh\delta+b_3\coth\delta)]^2]\cr
\bar M_{13} &= {\cR\over \r^2-2m\r}[-2m^2\sR\stg\cd+{1\over G_n}
[m^2\cd\sd(1\cr
&+\stg\stR)-m(m-\r)(b_2\tanh\delta+b_3\coth\delta)]\cr
&[2m^2\sg(s_r)-m(m-\r)b_1\tanh\gamma]]\cr
\bar M_{14} &= -{\sqrt{2}\over \r^2 -2m\r}[2m^2\sR\sg\cg\cd-{m(m-\r)\over
\sqrt\Delta}\cr
&(\sR\ca\sg\cd((s_rs_t)+\cb\cd+\ca\cg)\cr
&+(\cg(s_t)-\ca(s_r))(\cb\sd+(s_t)\sR\sg\cd)\cr
&+{G_p\over G_n}(m^2\cd\sd(1+\stg\stR)\cr
&+2m(m-\r)(b_2\tanh\delta+b_3\coth\delta)]\cr
\bar M_{15} &= {\sqrt{2}\over \r^2 -2m\r}[m^2\sR\sg\cg\cd+{m(m-\r)\over
\sqrt{\Delta}}\cr
&(\sR\ca\sg\cd((s_rs_t)+\cb\cd+\ca\cg)\cr
&+(\cg(s_t)-\ca(s_r))(\cb\sd+(s_t)\sR\sg\cd)\cr
&+{1\over G_n}m^2(a_1-\cg(s_r))(m^2\cd\sd(1+\stg\stR)\cr
&+2m(m-\r)(b_2\tanh\delta+b_3\coth\delta)]}}
\eqn\msII{\eqalign{\bar M_{16}&=1-{2m^2\ctR\stg\over 4G_n(\r^2-2m\r)}[4G_n
-(2m\sR\sg-(m^2-m\r){b_1\over\cg})^2]\cr
\bar M_{17} &= {\sqrt{2}\cR\sg\over \r^2-2m\r}[-m^2\cg+{m(m-\r)\over
\sqrt{\Delta}}(\cg(s_t)^2\cr
&\!\!\!\!\!\!\!+\ca (\cb\cd+\ca\cg))+{G_p\over 2G_n}(2m^2(s_r)
-m(m-\r){b_1\over\cg})]\cr
\bar M_{18} &= {\sqrt{2}\cR\sg\over \r^2-2m\r}[m^2\cg+{m(m-\r)\over
\sqrt{\Delta}}(\cg(s_t)^2\cr
&+\ca(\cb\cd+\ca\cg))+{1\over 2G_n}m^2(a_1-\cg(s_r))\cr
&(2m^2(s_r)-m(m-\r){b_1\over\cg})]}}
\eqn\newpsi{\eqalign{\Psi_1 &=\sqrt{2}[2m^2(b_4\sb\cb+b_5\coth\beta
-\sb\cd-(s_ts_r)\coth\beta)\cr
&(\r^3-m\r^2(3-b_{\Delta})+2m^2\r(1-b_{\Delta}))+m\r(\sb\cd+(s_ts_r)\coth\beta)
\cr &(\r^3-2m\r^2-m^2(\r-2m)(a_{\Delta}-2b_{\Delta})]\cr
\Psi_2 &= \sqrt{2}\cT\sa[m^2\r^3({b_1\over\cg}-4(s_r))+\r^2m^3
({b_1\over\cg}(b_{\Delta}-3)\cr
&+(s_r)(6-a_{\Delta})+\r 
m^4(2{b_1\over\cg}(1-b_{\Delta})+2(s_r)(a_{\Delta}-2))+m\r^4(s_r)]\cr
\Psi_3 &= \sqrt{2}[(b_2\tanh\delta+b_3\coth\delta)(\r^4m+2\r m^4
(a_{\Delta}-2)+m^3\r^2(6-a_{\Delta})\cr
&-4m^2\r^3)
+2\cd\sd(1+\stg\stR)(2m^4\r(1-b_{\Delta})\cr
&+m^3\r^2(b_{\Delta}-3)+m^2\r^3)]\cr
\Psi_4 &= \sqrt{2}\cR[m^2\r^3(2\sg(s_r)-2b_1\tanh\gamma)+\r^2m^3
(2\sg(s_r)(b_{\Delta}-3)\cr &+{b_1\over 2}\tanh\gamma(6-a_{\Delta})
+\r m^4(4\sg(s_r)(1-b_{\Delta})+b_1\tanh\gamma(a_{\Delta}-2))\cr
&+m\r^4{b_1\over 2}\tanh\gamma]\cr
\Psi_5&= m\r[b_1\r^3+2(a_1-2b_1+\cg(s_r))m\r^2+m^2\r(2b_{\Delta}
(a_1+\cg(s_r))\cr
&-b_1a_{\Delta}-6(a_1-b_1+\cg(s_r)))+2m^3(b_1a_{\Delta}+2(a_1-b_1+\cg(s_r))\cr
&-2b_{\Delta}(a_1+\cg(s_r))]\cr
\Psi_6 &= 2m^2\r(a_1-\cg(s_r))[\r^2-m\r(3-b_{\Delta})+2m^2(1-b_{\Delta})]}}
\eqn\allpqr{\eqalign{p_1 &=(\bar M_1-1)+{2(G_nM_1-G_p\psi_1)^2\over
(\r^2-2m\r)\Phi_dG_n}\cr 
&p_2 = \bar M_2+{2(G_nM_1-G_p\psi_1)
(G_nM_2-G_p\psi_2)\over (\r^2-2m\r)\Phi_dG_n}\cr
p_3 &= (\bar M_7-1)+{2(G_nM_2-G_p\psi_2)^2\over 
(\r^2-2m\r)\Phi_dG_n}\cr  
&q_1 = \bar M_3+{2(G_nM_1-G_p\psi_1)
(G_nM_3-G_p\psi_3)\over (\r^2-2m\r)\Phi_dG_n}\cr
q_2 = &\bar M_4+{2(G_nM_1-G_p\psi_1)
(G_nM_4-G_p\psi_4)\over (\r^2-2m\r)\Phi_dG_n}\cr
&q_3 = \bar M_8+{2(G_nM_3-G_p\psi_3)
(G_nM_2-G_p\psi_2)\over (\r^2-2m\r)\Phi_dG_n}\cr
q_4 &= \bar M_9+{2(G_nM_2-G_p\psi_2)
(G_nM_4-G_p\psi_4)\over (\r^2-2m\r)\Phi_dG_n}\cr
&r_1 = (\bar M_{12}-1)+{2(G_nM_{3}-G_p\psi_3)^2
\over (\r^2-2m\r)\Phi_dG_n}\cr
r_2 &= \bar M_{13}+{2(G_nM_3-G_p\psi_3)
(G_nM_4-G_p\psi_4)\over (\r^2-2m\r)\Phi_dG_n}\cr
&r_3 = (\bar M_{16}-1)+{2(G_nM_4-G_p\psi_4)^2
\over (\r^2-2m\r)\Phi_dG_n}}}

\vfil\eject
\listrefs
\bye